\documentstyle[12pt]{article}

\makeatletter
\if@twoside
\else \oddsidemargin 00pt \evensidemargin 00pt
\fi
\let\@eqnsel = \hfil

\expandafter\ifx\csname mathrm\endcsname\relax\def\mathrm#1{{\rm #1}}\fi
\@ifundefined{mathrm}{\def\mathrm#1{{\rm #1}}}{\relax}

\makeatother

\def\gsim{\ \rlap{\raise 2pt \hbox{$>$}}{\lower 2pt \hbox{$\sim$}}\ }
\def\lsim{\ \rlap{\raise 2pt \hbox{$<$}}{\lower 2pt \hbox{$\sim$}}\ }

\def\ea{{et al.}}

\def\npb#1{Nucl. Phys. {\bf B#1}}
\def\plb#1{Phys. Lett. {\bf B#1}}
\def\prd#1{Phys. Rev. {\bf D#1}}
\def\prl#1{Phys. Rev. Lett. {\bf #1}}

\def\zpc#1{Z. Phys. {\bf C#1}}

\def\hb{\hfill\break}

\begin{document}
\thispagestyle{empty}

\rightline{CERN-TH.7443/94}
\rightline{UM-TH-94-34}

\rightline{hep-ph/9409310}

\vskip 0.5truecm

\begin{center}
{\Large \bf      NEW NEUTRAL GAUGE BOSONS AND NEW HEAVY FERMIONS
IN THE LIGHT OF THE NEW LEP DATA
\par} \vskip 2.em

{\large		
{{\sc Enrico Nardi}$^{a,\dagger}$, {\sc Esteban Roulet }$^{b}$ and
{\sc Daniele Tommasini }$^{b}$}
\\[1ex] 
{$^{a}$\it Randall Laboratory of Physics, University of
Michigan}\\
{\it Ann Arbor, MI 48109--1120, U.S.A.}\\[2ex]
\vskip .3cm
{$^{b}$\it Theory Division, CERN, CH-1211, Geneva 23, Switzerland}\\
\par}
\end{center} \par
\bigskip
\centerline{\bf Abstract} \par
\vskip 2truemm
\noindent
We derive limits on a class of new physics effects that are naturally
present in grand unified theories based on extended gauge groups, and
in particular in $E_6$ and $SO(10)$ models. We concentrate on $i$) the
effects of the mixing of new neutral gauge bosons with the standard
$Z_0$; $ii$) the effects of a mixing of the known fermions with new
heavy states. We perform a global analysis including all the LEP data
on the $Z$ decay widths and asymmetries collected until 1993, the SLC
measurement of the left--right asymmetry, the measurement of the $W$
boson mass, various charged current constraints, and the low energy
neutral current experiments. We use a top mass value in the range
announced by CDF. We derive limits on the $Z_0$--$Z_1$ mixing, which are
always $\lsim 0.01$ and are at the level of a few {\it per mille} if
some specific model is assumed. Model-dependent theoretical relations
between the mixing and the mass of the new gauge boson in most cases
require $M_{Z'} > 1\,$TeV. Limits on light--heavy fermion mixings are
also largely improved with respect to previous analyses, and are
particularly relevant for a class of models that we discuss.
\par
\begin{center}
Revised version (november 1994), to appear in Phys. Lett. B
\end{center}
\vskip .5truecm
\bigskip
\noindent
--------------------------------------------\phantom{-} \hb
\noindent
$\dagger$ Address from November 1994 :
{\it Department of Particle Physics, Weizmann Institute of
Science,}\ P.O.B. 26, Rehovot, 76100 Israel. \hb
\bigskip
\noindent
E-mail addresses: \hb
nardi@wiswic.weizmann.ac.il,
roulet@vxcern.cern.ch,tommasini@vxcern.cern.ch
\vskip .5cm
\noindent CERN-TH.7443/94 \par
\noindent UM-TH-94-34 \par
\vskip .15mm
\noindent September 1994 \par
\vfill\eject

\setcounter{page}{1}
\null

\phantom{\cite{ll}\cite{fit}\cite{mixlep}\cite{fit6}\cite{zp-new}
\cite{cdf-mtop} }

\section{Introduction}

The sensitivity of LEP experiments to the direct production of new
particles has not increased significantly with respect to that
achieved after the first-year runs. However, the accumulation of large
statistics and the improvements on the systematics now allows not only
to test with much more detail the predictions and consistency of the
standard model (SM),
including the virtual effects of the top quark
and Higgs boson, but also to improve considerably the ability to
search for (or constrain) some subtle indirect manifestations of new
physics beyond the SM.

Among these last ones, the implications of the combined LEP
measurements up to 1993 are of particular importance for new neutral
gauge bosons that could mix with the standard $Z_0$ (so that the $Z$ boson
mass eigenstate has a small component with non-standard couplings) and
also for heavy fermions mixed with the known ones. In fact if the new
fermions have non-canonical $SU(2)\times U(1)$ quantum numbers (e.g.
left-handed singlets or right-handed doublets) they modify the
couplings of the electroweak gauge bosons with the light particles.

These new kinds of physics are a common feature of many GUT theories,
such as $SO(10)$ and E$_6$. The search for the tiny effects mentioned
above then allows us to look indirectly for the new states
predicted by these models,  even if their direct production
is unaccessible at the energies
achievable with present colliders. Global
constraints on these effects have been regularly
performed in the past using the available electroweak data,
\cite{ll}--\cite{zp-new}. In
this paper we show that the inclusion of LEP and SLC data up to 1993
allows a significant improvement of the constraints on the deviations of the
fermion couplings with respect to their SM values and hence strengthen
the bounds on the above-mentioned mixings,  in some cases even by an
order of magnitude. The value of the top quark mass recently
announced by the CDF collaboration \cite{cdf-mtop}, $m_t=174\pm
10^{+13}_{-12}$,
is also relevant for this analysis, since some bounds (such
as those on $Z_0$ mixing with an additional gauge boson or those on the
mixing of the $b$ quark) are correlated with it.

Finally we briefly discuss whether it is possible
that new physics effects of the kind discussed here could
account for the deviation  from the SM expectations of
some measurements,  such as
$\Gamma_b^{LEP}$, $A_{LR}^{SLC}$ and $A^{FB}_\tau$.
We can anticipate that we find essentially negative results.

\section{$Z_0$--$Z_1$ mixing}

The formalism describing the mixing of the standard neutral $Z_0$
boson of the electroweak gauge group
${\cal G}_{SM}=SU(2)\times U(1)$ with a new $Z_1$ associated
with an extra $U'(1)$ factor has been discussed at length in the past
\cite{fit6,zp-new}. Here we just recall a few relevant points.

In order to span a wide range of $Z'$ models, we will as usual take
the $U'(1)$ as a combination of the two additional Abelian factors in
the decomposition $E_6\to SO(10)\times U(1)_\psi\to SU(5)\times
U(1)_\chi\times U(1)_\psi$, where ${\cal G}_{SM}$ is assumed to
be embedded
in the $SU(5)$ factor.  We hence parametrize the new gauge boson
as
\begin{equation}
Z_1=s_\beta Z_\psi+c_\beta Z_\chi,
\end{equation}
 where $s_\beta \equiv \sin
\beta$, $c_\beta \equiv \cos\beta$. We will present results for the most
commonly considered $\chi$, $\psi$ and $\eta$ models, corresponding
respectively to $s_\beta=0$, 1 and $-\sqrt{5/8}$.

A mixing between $Z_0$ and $Z_1$ leads us to the two mass eigenstates
\begin{equation}
\left(\begin{array}{c}Z\\Z'\end{array}\right)=
\left(\begin{array}{cc}c_\phi&s_\phi\\-s_\phi&c_\phi\end{array}\right)
\left(\begin{array}{c}Z_0\\Z_1\end{array}\right).
\end{equation}

Although one may consider
$\phi$ as being a free parameter, one should remember that in
any given model one generally has $\phi\simeq CM_Z^2/M_{Z'}^2$, where
$C\sim O(1)$ is fixed once the vacuum expectation values
(VEVs) of the Higgs fields giving
masses to the gauge bosons are specified. This theoretical relation
between $M_{Z'}$ and $\phi$ has the important implication
that the very stringent constraints on the mixing angle $\phi$
obtained by LEP at the $Z$-pole (see below) induce, once a model
fixing $C$ is assumed, an indirect bound on $M_{Z'}$ typically much
stronger ($M_{Z'}\gsim$ 1 TeV) than those arising from direct $Z'$
searches at the Tevatron ($M_{Z'}\gsim 450$ GeV for 25 $pb^{-1}$
of integrated luminosity
\cite{zp-paco}) or those resulting
from the effects of $Z'$ exchange on low-energy neutral current
experiments ($M_{Z'}\geq 200-300$ GeV)
\cite{fit6,zp-new,zp-paco}.
In view of these bounds we will neglect
in the following $Z'$ exchange and $Z$--$Z'$ interference
effects in the neutral current (NC) processes,
and we will only consider the modifications of the
$Z$ couplings to fermions induced by the small
admixture with the $Z_1$.

Due to the $Z_0$--$Z_1$ mixing,
the vector and axial-vector fermion couplings
appearing  in  the NC
$J_Z^\mu=\bar\Psi^f(v^f+a^f\gamma_5)\gamma_\mu\Psi^f$,
which couples to the physical $Z$ boson, read\footnote{The sine of the weak
mixing angle $s_W$ appears due to the normalization of the $U'(1)$ coupling
\cite{fit6}.}
\begin{eqnarray}
v^f&=&c_\phi v^f_0+s_\phi s_Wv^f_1,\\
a^f&=&-s_\phi a^f_0+c_\phi s_Wa^f_1.
\end{eqnarray}
Within the SM, and including radiative corrections, one has
\begin{equation}
v^f_0=\sqrt{\rho_f}\, [t_3(f_L)-2\,Q^f \sin^2\theta_{eff}^f]~~~~,~~~~
a^f_0=\sqrt{\rho_f}\, t_3(f_L),
\end{equation}
where $\sin^2\theta_{eff}^f$ and the $\rho_f$ factors
have  been evaluated by means
of the ZFITTER code\footnote{We thank D. Bardin for
providing us with the 1994 updated version of the program.} \cite{zfitter},
as functions of the input parameters $m_t$, $\alpha_s(M_Z)$ and $m_H$.
The $Z_1$
couplings $v_1$ and $a_1$ depend on the assumed $U'(1)$ model (i.e. on
$s_\beta$) and can be found in refs. \cite{fit6,zp-new}.
The effects of the SM radiative correction
induced by the mixings with the new particles,
as well as the
radiative effects of new physics, are expected to be small and
have been neglected. A more detailed justification of
this assumption can be found in \cite{fit6}.

Since we are  neglecting $Z^\prime$ propagator effects, the
only quantity in which the $Z'$ mass appears explicitly is
$\rho_{mix}=1+(M_{Z'}^2/M_Z^2-1)s_\phi^2$.
This term affects the $SU(2)$ gauge coupling
deduced using as numerical inputs  $G_F$,
$\alpha$ and the value of $M_Z$ measured at LEP, thus
modifying both the overall strengths $\rho_f$ and the
$\sin^2\theta_{eff}^f$ factors. Since the effects of
$\rho_{mix}$ in the LEP observables are crucial to constrain the mixing
$\phi$,
the limits on the $Z_0$--$Z_1$ mixture will depend on the
$Z'$ mass, generally improving with larger $M_{Z'}$ values.

A second remark is that $\rho_{mix}$
enters as a multiplicative factor in the effective $\rho$ parameter.
Then the combined appearance of $\rho_{mix} \cdot \rho_{top}$,
with $\rho_{top}\simeq 1+{3G_Fm_t^2\over 8\sqrt{2}\pi^2}$, induces
a strong
correlation between the gauge boson mixing and the top mass.
Hence the top mass measurement by CDF \cite{cdf-mtop}
turns out to be relevant to
establish precise bounds on the mixing angle $\phi$.

\section{Fermion mixing}

A mixture of the known fermions with new heavy states can in general
induce both flavour changing (FC) and non-universal flavour diagonal
vertices among the light states. The first ones are severely
constrained (for most of the charged fermions) by the limits on rare
processes \cite{fc-limits}. Here we aim to constrain the second ones
by means of the large set of precise electroweak data.

Due to the extremely tight constraints on the FC
mixings \cite{fc-limits},
neglecting them will not affect our numerical analysis
on the flavour diagonal ones,
since in general the limits on the latter ones turn out to be larger
by some orders of magnitude.
 From a theoretical point of view,
the absence of FC parameters in the formalism that we will
outline here is equivalent  to the assumption that
different light mass eigenstates have no
mixtures with the same new state \cite{ll}.

The couplings
of the light charged fermions can then be described with just two
parameters for each flavour: $(s^f_\alpha)^2\equiv
\sin^2\theta^f_\alpha$, $\alpha=L,R$, which account for the mixing
with exotic states (i.e. having non-canonical $SU(2)\times
U(1)$ quantum numbers) of each of the two fermion chiralities.
Since the mixing always involves states of equal electric charges,
only the piece proportional to the weak isospin $t_3(f)$ in (5) is
affected by the fermionic mixing. In particular, the chiral couplings
$\epsilon_{L,R}^f=(v^f\pm a^f)/2$ are modified according to (see eq.
2.15 of ref. \cite{fit6})
\begin{equation}
\epsilon_\alpha^f=t_3(f_\alpha)-Q^f\sin^2\theta^f_{eff}+\left[
t_3(f^{\cal N}_\alpha)-t_3(f_\alpha)\right] (s^f_\alpha)^2~~,
{}~\alpha=L,R~~,
\end{equation}
where $t_3(f_\alpha^{\cal N})$ is the isospin of the new state
$f^{\cal N}$ that mixes with the known state $f$.
(For notational simplicity we omit hereafter the $\sqrt{\rho_f}$ factors in
the expressions for the couplings.)
Eq. (6) shows that when a doublet state is mixed with a singlet,
the isospin-dependent  part of the coupling
is reduced by a factor $(c^f_L)^2$, while the mixing of a
singlet ($t_3(f_R)=0$) with a new exotic doublet ($t_3(f^{\cal
N}_R)=\pm 1/2$) induces a coupling
proportional to $t_3(f_R^{\cal N})(s^f_R)^2$. Clearly, a mixing
between
states of the same isospin does not affect the overall electroweak
couplings.
Here we will only consider
mixings with new states that are either exotic singlets or
exotic doublets, i.e. $t_3(f_L^{\cal N})=t_3(f_R)=0$ and
$t_3(f_R^{\cal N})=t_3(f_L)=\pm 1/2$. Then,
in the absence of extra new gauge bosons, we have
\begin{eqnarray}
v^f&=&t_3(f_L)[1-(s^f_L)^2+(s^f_R)^2] -2Q^f\sin^2\theta_{eff}^f\\
a^f&=&t_3(f_L)[1-(s^f_L)^2-(s^f_R)^2].
\end{eqnarray}
\noindent
The mixing among the neutral fermionic states is not so simple, both
because of the lack of strong evidence against FCNC among neutrinos and
because of the possible existence of more than one type of exotic
states (singlets, exotic doublets with $t_3(N_L)=-1/2$, etc.
\cite{ll,fitnu}).
However, after summing over the undetected final neutrinos
and neglecting $O(s^4)$ terms, the
different NC observables can be obtained
by replacing the neutrino
couplings in the SM expressions by effective couplings,
which  depend on just one mixing angle for each flavour:
\begin{equation}
v_{\nu_i}=a_{\nu_i}={1\over 2}-{\Lambda_i\over 4}(s^{\nu_i}_L)^2.
\end{equation}
The additional parameter $\Lambda$ describes the type of state
involved in the mixing and, for instance,
for a mixing with new ordinary, singlet or exotic doublet neutrinos
we have $\Lambda=0$, 2 or 4 respectively.

An important indirect effect of the presence of new fermions is to
alter the prediction for $\mu$ decay, in such a way that the effective
$\mu$-decay constant $G_\mu = 1.16637(2) \times 10^{-5}$ GeV$^{-2}$ is
related to the fundamental coupling $G_F$ through the fermion mixing
angles \cite{ll,fit},
\begin{equation}
G_\mu = G_F c_L^{e}c_L^{\mu}c_L^{\nu_e}c_L^{\nu_\mu}.
\label{eq:fitnu9}
\end{equation}
As a consequence, all the observables that depend
on the strength of the weak interactions $G_F$ are affected by
the mixing angles $\theta_L^{e}$, $\theta_L^{\mu}$,
$\theta_L^{\nu_e}$ and $\theta_L^{\nu_\mu}$.
This is the case, for instance, for the $W$ boson mass, for the effective
couplings of the fermions with the $Z$ boson, and
for the Cabibbo--Kobayashi--Maskawa (CKM) matrix elements
\cite{ll,fit}.

The complete formalism describing fermion mixings and
also the simultaneous
presence of $Z_0$--$Z_1$ mixing is given in ref. \cite{fit6}.

\section{Theoretical expectations for the fermion mixings}

As regards the theoretical expectations for the
mixing of the known fermions with new heavy states,
there is  no exact model-independent relation
between the masses of the heavy partners and the
corresponding mixings.
However, in the framework of some classes of models,
it is still possible to make some general statements
and/or work out some order-of-magnitude estimates for the
mixings.

For the charged states, the L (or R) mixing angles result
from the diagonalization of the $N\times N$ symmetric
squared mass matrix for the known and the new states
${\cal M}{\cal M}^\dagger$ ( or ${\cal M}^\dagger{\cal M}$ ).
We know that the relevant eigenvalues must satisfy the hierarchy
$m^2_{\rm light} \ll m^2_{\rm heavy}$ (with $m_{\rm heavy}\gsim 100$
GeV),
and we can outline two main mechanisms that would
naturally produce
such a pattern for the light and heavy masses.

\noindent
{\it a)}\quad {\it See-saw models} \hb
In these models the general form of the squared mass matrix is
\begin{equation}
{\cal M}{\cal M}^\dagger \sim
\left(\begin{array}{cc}\delta^2 &d^2  \\
d^2 &\sigma^2 \end{array}\right),
\label{mass-matrix}
\end{equation}
with $\delta$, $d \ll \sigma$.
If $\delta \sim d$,
as is the case if both these entries are
generated by VEVs of standard Higgs doublets,
we expect for the mass eigenvalues
$m_{\rm light} \sim \delta$, $m_{\rm heavy}\sim \sigma$, and
$s_{L,R} \sim d^2/\sigma^2 \sim
m^2_{\rm light} / m^2_{\rm heavy}$.
A different scenario appears when $\delta \lsim  d^2/\sigma$, for
which $m_{\rm light} \sim d^ 2 /\sigma$, $m_{\rm heavy}\sim \sigma$,
and  $s_{L,R} \sim d^2/\sigma^2 \sim m_{\rm light} / m_{\rm heavy}$.
 Assuming $m_{\rm heavy}\gsim 100\,$GeV,
we see that in the Dirac see-saw case
the expectations for the
mixings are quite small. In the most favourable case of the bottom
quark mixing, it can be as large as $(s^b_{L,R})^2 \sim
2\times 10^{-3}$, which
turns out to be at the limit of the present experimental sensitivity.

\noindent
{\it b)}\quad {\it Quasi-degenerate mass matrices} \hb
It can happen that, as a consequence of some symmetries,
in first approximation
the light--heavy fermion mass matrices are degenerate.
This implies that even if all the entries in
the mass matrices are large, some states (corresponding to the
light fermions) are massless, and would
acquire tiny masses due to
small flavour-dependent perturbations.
To give a simple example of this mechanism,
let us introduce a vector-like singlet of new fermions $F_L$ and $F_R$,
of the same charge and colour quantum numbers as those of the
$f_L$ component of a standard  electroweak doublet, and of the
corresponding electroweak singlet $f_R$.
The general mass term reads
\begin{equation}
{\cal L}_{mass} =
\lambda_0  \overline {F_L} F_R S +
\lambda_1  \overline {F_L} f_R S +
\gamma_0  \overline {f_L} F_R  D +
\gamma_1  \overline {f_L} f_R  D ,
\label{l-mass}
\end{equation}
where $S$ and $D$ are respectively a singlet and a doublet VEV.
Let us also assume that because of some symmetries,
in first approximation
$\lambda_0 \simeq \lambda_1$ and $\gamma_0  \simeq \gamma_1$,
and let us absorb these Yukawas in the $D$ and $S$ VEVs.
Then, the light--heavy mass matrix squared that determines
the ordinary--exotic L mixing angle reads
\begin{equation}
{\cal M}{\cal M}^\dagger \sim
2\, \left(\begin{array}{cc}D^2 &D S \\
D S &S^2 \end{array}\right),
\label{deg-mass-matrix}
\end{equation}
and is clearly degenerate,
implying
$m_{\rm light} \simeq 0$ up to perturbations. At the same time,
the ordinary--exotic L mixings are expected to be large, and
could even be close to maximal.
The expectations for the neutral sector were
described in \cite{fitnu}, where it was shown that a
similar mechanism can also generate large light--heavy mixings even for
massless neutrinos.

Clearly, in contrast to the see-saw case,
models of this kind can be effectively constrained by
analysing the most precise electroweak data, and in fact the tight
bounds that we will derive for some mixings tend
to disfavour this mechanism
for the generation of the known fermion masses.

\section{Experimental constraints}

Within the SM, the precise electroweak experiments allow to constrain
the values of the input parameters $m_t$, $\alpha_s(M_Z)$ and $m_H$,
and  an overall satisfactory agreement is found whith the predictions
for a heavy top mass \cite{pietrzyk}, compatible with the range
obtained by CDF. For instance, for $m_t=170$ ~GeV, $\alpha_s=0.12$ and
keeping hereafter the Higgs mass fixed at $m_H=250$~GeV, for most
observables the measured value is actually very close to the
theoretical predictions, making the total $\chi^2$ per degree of
freedom reasonably low ($<2$). However, there are a few exceptions for
which recent data show some noticeable disagreement with respect to
the SM expectations. A well-known case is the SLC measurement of the
left--right polarized asymmetry $A_{LR}$ \cite{alr-slc} ($\chi^2\sim
10$ for the above mentioned choice of input parameters). Some LEP
results also show sizeable deviations. This is the case for the ratio
of the $Z$ width into $b$ quarks to the total hadronic width,
$R_b\equiv \Gamma_b/\Gamma_h$ ($\chi^2\sim4.5$), and for the $\tau$
forward--backward asymmetry $A^{FB}_\tau$ ($\chi^2\sim7$)
\cite{lep94}. (Clearly the actual value of the $\chi^2$ function
depends on the values adopted for the input parameters.)

For our analysis we have used the CC constraints on
lepton universality and on CKM unitarity, the $W$ mass measurement,
as well as the NC constraints from the LEP and SLC measurements
at the $Z$ peak.

The best test of $e$--$\mu$ universality comes
from $\pi\to e\nu$ compared to $\pi\to \mu\nu$.
The ratio of the electron to the muon couplings to the $W$ boson,
extracted from the TRIUMF \cite{triumf92} and PSI \cite{psi92}
measurements, is
$
\left(g_e/ g_\mu\right)^2=0.9966\pm0.0030
$ \cite{fitnu}.

Universality among the $\mu$ and $\tau$ leptons is tested by
the $\tau$ leptonic decays compared to $\mu$ decay, giving
$(g_\tau/g_\mu)^2=0.989\pm0.016$ \cite{roney}.
A second test comes from $\tau\to\pi(K)\nu_\tau$,
which gives $(g_\tau/g_\mu)^2=1.051\pm0.029$ \cite{roney};
this is almost $2\sigma$ off the SM,
and hardly compatible with the above determination from $\tau$ decays.
The use of this determination affects mainly our bounds for the mixing of
the $\tau$ neutrino with new ordinary states, as discussed in Ref.
\cite{fitnu}.

For the test of the unitarity of the first row of the CKM matrix,
we use the determination
$\sum_{i=1}^3|V_{ui}|^2 = 0.9992\pm 0.0014 $
of Ref. \cite{sirlin93}, and for the $W$ mass we take the average
$M_W=80.23\pm0.18$ \cite{pietrzyk} of the CDF and UA2 experimental
values.

For the $Z$-peak data, we have included
the measurements of the total $Z$ width
$\Gamma_Z$, the hadronic peak cross section $\sigma_h^0$, the ratios
$R_e$, $R_\mu$, $R_\tau$ of the total hadronic width to the
flavour-dependent leptonic ones,
the bottom and charm ratios $R_b$ and $R_c$
and forward--backward asymmetries $A^{FB}_b$ and $A^{FB}_c$,
and the leptonic unpolarized asymmetries
$A^{FB}_e$, $A^{FB}_\mu$ and $A^{FB}_\tau$. All the data up to 1993 as
well as all the relevant experimental correlations given in Ref.
\cite{lep94} have been taken into account in our analysis.
We also include in our set of constraints the measurements of the
left--right polarization asymmetry at SLC, $A_{LR}=0.1637\pm0.0075$
\cite{alr-slc}, and the measurement of the ``theoretically equivalent"
quantity
$A_e^0={2 a_e v_e\over a_e^2 + v_e^2}=
0.120\pm0.012$ which has been inferred by the LEP collaborations from
the angular distribution of the $\tau$ decay products \cite{lep94}.
These two different determinations of the same theoretical quantity
are both more than $2 \sigma$ off the
SM value ($A^0_e=0.1419$ for our set of input parameters) and are in even
more serious conflict between them, possibly
indicating some problem in the analysis of the experimental data or an
unlucky fluctuation.

We always use
values for the observables that are extracted from the data without
assuming universality, which is expected to be violated by the fermion
mixings in the models we are considering. It is interesting to notice that,
while the experimental leptonic partial width of the $Z$ boson are in good
agreement with the hypothesis of universality, some hint of a
discrepancy may be present
in the fitted flavour-dependent forward--backward asymmetries, which
are $A^{FB}_e=0.0158\pm0.0035$, $A^{FB}_\mu=0.0144\pm 0.0021$ and
$A^{FB}_\tau=0.0221\pm 0.0027$ \cite{lep94,pietrzyk}.

Finally, we have also included in our data set the (updated)
low-energy NC constraints (deep inelastic $\nu$ scattering and atomic
parity violation). Although less effective than the
$Z$ peak data for constraining the kind of physics we are considering,
they turn out to be relevant for our analysis in
the case of the `joint' fits to be discussed below.

\section{Results}

After constructing a $\chi^2$ function with all
the experimental measurements discussed in the previous section,
we have derived bounds on the mixing parameters
by means of the MINUIT package.

\begin{table}[p]
\begin{center}
\caption{90\% c.l. lower ($\phi_-$) and upper ($\phi_+$) bounds on the
the $Z-Z'$ mixing angle
$\phi$, in units of $10^{-2}$, for the $\psi$, $\eta$ and $\chi$ models.
The limits correspond to different values
of the top mass $m_t$ and the strong coupling constant $\alpha_s$, with the
Higgs mass fixed to $m_H=250$ GeV. They have been obtained by chosing the
most conservative values as
$M_{Z'}$ is allowed to vary from $\sim500$ GeV to infinity.}
\vskip 1cm
\begin{tabular}{|c|c|c|c|c|c|}
\hline
$m_t$ [GeV] & $\alpha_s$ & $E_6$ model
              & $\phi_- \, [10^{-2}]$ & $\phi_+\, [10^{-2}]$\\
\hline
\hline
150 & 0.11 & $\psi$ & 0   & 1.1 \\
    &      & $\eta$ & 0   & 1.3 \\
    &      & $\chi$ & 0   & 0.75 \\
\hline
    & 0.12 & $\psi$ & 0   & 0.81\\
    &      & $\eta$ & 0   & 1.0 \\
    &      & $\chi$ & -0.31& 0.43\\
\hline
    & 0.13 & $\psi$ & -1.0 & 0\\
    &      & $\eta$ & -1.0 & 0 \\
    &      & $\chi$ & -0.70& 0\\
\hline
200 & 0.11 & $\psi$ & -0.03& 0.57 \\
    &      & $\eta$ & -0.19& 0.60 \\
    &      & $\chi$ & 0   & 0.48 \\
\hline
    & 0.12 & $\psi$ & -0.28& 0.32\\
    &      & $\eta$ & -0.33& 0.39\\
    &      & $\chi$ & -0.18& 0.25\\
\hline
    & 0.13 & $\psi$ & -0.43 & 0.11\\
    &      & $\eta$ & -0.38 & 0.23\\
    &      & $\chi$ & -0.38& 0.06\\
\hline
\end{tabular}
\end{center}
\end{table}

Regarding the gauge boson mixing $\phi$, we give for the
 {\it unconstrained models} (e.g. with $M_{Z'}$ independent of
$\phi$) conservative bounds
obtained letting the $Z'$ mass to take values in the range
$M_{Z'}>500$~GeV and taking the extreme values $\phi_\pm$ that
remain allowed at 90\% c.l.. In this way we obtain
$$-0.0056<\phi<0.0055 \ \ \ \ \  (\psi\ \mbox{model})$$
\begin{equation}
-0.0087<\phi<0.0075 \ \ \ \ \  (\eta\ \mbox{model})\end{equation}
$$-0.0032<\phi<0.0031 \ \ \ \ \  (\chi\ \mbox{model})$$
These results have been obtained chosing for the input parameters the
values
$m_t=170$ GeV, $m_H=250$ GeV and $\alpha_s=0.12$, which provide
a good agreement between the experimental observables and
the SM predictions (corresponding to vanishing $Z$-$Z'$
and fermion mixings).
Since the bounds on $\phi$ depend on the choice of input parameters,
we show in Table 1 how the constraints are modified for $m_t=150$ and
200 GeV and for $\alpha_s=0.11$, 0.12 and 0.13\footnote{For
a detailed discussion of the $m_t$ (and $m_H$)
dependence, see the last two references in  \cite{zp-new}.}.
It is apparent that
the bounds become tighter for increasing $m_t$.
This can be easily traced back to the fact that
larger (absolute) values of $\phi$ and of $m_t$ both tend to increase the
value of the effective $\rho$ parameter
$\sim \rho_{mix} \cdot \rho_{top}$.
For this reason the CDF lower limit on
$m_t$ is relevant for constraining $\phi$.
On the other hand, in the models considered here,
increasing values of $\alpha_s$ lead to a shift towards negative
$\phi$ values of the allowed region.

The previous bounds get also somewhat relaxed if one
allows for the simultaneous presence of the fermion mixings,
which can produce
compensating effects. In this case, keeping from now on the same choice
($m_t=170$ GeV, $m_H=250$ GeV, $\alpha_s=0.12$) for the input parameters,
we get the 90\% c.l. constraints
$$-0.0066<\phi<0.0071\ \ \ \ (\psi\ \mbox{model}),$$
\begin{equation}
-0.0087<\phi<0.010\ \ \ \  (\eta\ \mbox{model}),
\end{equation}
$$-0.0032<\phi<0.0079\ \ \ \  (\chi\
\mbox{model}).$$

In contrast, tighter bounds result if one considers {\it constrained
models}, that is assuming a relation between the gauge boson mixing and
$M_{Z'}$ of the form  $\phi\simeq C M_Z^2/M_{Z'}^2$, where $C$ can be
evaluated once the Higgs sector is specified.
In this case the bounds on $\phi$ translate also into indirect
constraints on $M_{Z'}$.
The following results have been derived
by assuming for each model  a {\it minimal} Higgs content and the absence
of singlet VEVs.
For the $\psi$ model, denoting by $\sigma\equiv (v_u/v_d)^2$
the square of the ratio of the scalar VEVs giving masses respectively
to the $u$ and $d$-type quarks, we have
$C=-{\sqrt{10}\over 3}s_W
{\sigma-1\over \sigma+1}$.
For $\sigma\to\infty$ we obtain $0\geq\phi>-0.0042$, which
implies the indirect constraint $M_{Z'}>1.0$
TeV, while, for instance, for $\sigma=2$ we obtain $0\geq\phi>-0.0052$,
corresponding to $M_{Z'}>0.52$ TeV.
For the $\eta$ model ($C={4\over 3}s_W
{\sigma-1/4\over \sigma+1}$) the bound
for $\sigma\to\infty$ is $0\leq\phi<0.0035$, implying $M_{Z'}>1.2$
TeV, while
for $\sigma=2$ we obtain $0\leq\phi<0.0054$, implying $M_{Z'}>0.76$
TeV.
We recall that the $Z_\chi$ of
the $\chi$ model is equivalent to the $Z'$ present in
$SO(10)$, being the two models different only with respect
to the fermion and scalar representations.
For the minimal Higgs content of $SO(10)$
($C=s_W\sqrt{2/3}$ \cite{lang-luo})
we obtain the constraint
$0\leq\phi<0.0028$, which implies
$M_{Z'}>1.2$ TeV for a $Z'$ from $SO(10)$.

\begin{table}[p]
\begin{center}
\caption{The 90\% c.l. upper bound on the
ordinary--exotic fermion mixing
parameters. The  `single' limits in the first column are obtained when
the remaining mixing parameters are set to zero. For the `joint' bounds
in the second column, cancellations among the effects of all the different
possible fermion mixings are allowed.
The third column gives the `joint' bound in
the $\chi$ model, taking into account the possible cancellations among
the effects of
all the ordinary--exotic mixing parameters present in $E_6$
as well as
of a $Z_0-Z_\chi$ mixing.
All the results presented correspond to the value $\Lambda
=2$ of the parameter describing the type
of new neutrinos involved in the mixing, with the fixed values
$m_t=170$ GeV, $m_H=250$ GeV and $\alpha_s=0.12$. }
\vskip 1cm
\begin{tabular}{|c||c|c|c|c|c|}
\hline
 & Single limit & Joint limit & $\chi$ model \\
\hline
\hline
$(s_L^e)^2$ & 0.0018 & 0.0065 &  \\
\hline
$(s_R^e)^2$ & 0.0020 & 0.0020 & 0.0024  \\
\hline
$(s_L^\mu)^2$ & 0.0017 & 0.0076 &  \\
\hline
$(s_R^\mu)^2$ & 0.0034 & 0.0059 & 0.0045 \\
\hline
$(s_L^\tau)^2$ & 0.0016 & 0.0058 &  \\
\hline
$(s_R^\tau)^2$ & 0.0030 & 0.0055 & 0.0037 \\
\hline
$(s_L^u)^2$ & 0.0024 & 0.012 &  \\
\hline
$(s_R^u)^2$ & 0.0090 & 0.015 &  \\
\hline
$(s_L^d)^2$ & 0.0023 & 0.013 & 0.0064 \\
\hline
$(s_R^d)^2$ & 0.019 & 0.029 &  \\
\hline
$(s_L^s)^2$ & 0.0036 & 0.0087 & 0.019 \\
\hline
$(s_R^s)^2$ & 0.021 & 0.060 & \\
\hline
$(s_L^c)^2$ & 0.0042 & 0.019 & \\
\hline
$(s_R^c)^2$ & 0.010 & 0.17 & \\
\hline
$(s_L^b)^2$ & 0.0020 & 0.0025 & 0.0045 \\
\hline
$(s_R^b)^2$ & 0.010 & 0.015 & \\
\hline
$(s_L^{\nu_e})^2$ & 0.0050 & 0.0066 & 0.0064 \\
\hline
$(s_L^{\nu_\mu})^2$ & 0.0018 & 0.0060 & 0.0046 \\
\hline
$(s_L^{\nu_\tau})^2$ & 0.0096 & 0.018 & 0.017 \\
\hline
\end{tabular}
\end{center}
\end{table}

Turning now to the fermion mixings, we have listed in table 2 the
updated 90\% c.l. bounds
obtained by allowing just one mixing to be present
(single bounds) or allowing for the simultaneous presence of all types
of fermion mixings (joint bounds). In the last case the constraints
are generally relaxed due to possible accidental cancellations among
different mixings. The bounds on the fermion mixings that can appear
in $E_6$ models are given in the third column. In this case we have
also allowed for the presence of mixing among the gauge bosons, which
somewhat relaxes the limits. We present the results obtained in the
$\chi$ model with the $Z_0$--$Z_\chi$ mixing as an additional free
parameter.

The constraints we have listed in table 2 correspond to the
particular value $\Lambda=2$. However we stress that only the bound on
$s^{\nu_\tau}_L$ depends significantly on the adopted value of
$\Lambda$, since the $\nu_e$ and $\nu_\mu$ mixings are mainly
constrained by CC observables, which do not depend on this parameter.
The LEP data alone already imply
$(s^{\nu_\tau}_L)^2<0.002/\Lambda_\tau$, which, due to the improvement
in the determination of the invisible width, is significantly better
than what obtained in previous analyses.  For $\Lambda_\tau\simeq 0$
the constraint on $s_L^{\nu_\tau}$ arises from CC observables and can
be found in ref. \cite{fitnu}.

The results in table 2 were obtained for the reference values
$m_t=170$ GeV and $\alpha_s=0.12$.
Allowing $m_t$ to vary in the range 150
to 200 GeV does not affect significantly the constraints
on the fermion mixings. In contrast,
increasing $\alpha_s$ up to  $\alpha_s=0.13$
worsens the limits on
some of the hadronic mixings
up to a factor $\sim 2$.

Besides strengthening the bounds on the new physics, one may also
wonder whether it could be possible to account for some of the
deviations with respect to the SM predictions that we have
mentioned previously, by means of the new physics effects that we
have been discussing here.
Regarding the $\sim 2 \sigma$ excess reported in the measurement of
$R_b$, the observed deviation ($\Gamma_b^{exp} > \Gamma_b^{SM}$)
has the opposite sign than the one
resulting from a mixing of the bottom quark with exotic states. In fact,
since
$\Gamma_b\propto v_b^2+a_b^2$,  at $O(s_{L,R}^2)$ we have
\begin{equation}
{\Gamma_b\over\Gamma_b^{SM}}\simeq 1 +
(s_L^b)^2{v_0^b+a_0^b\over (v_0^b)^2 + (a_0^b)^2} +
(s_R^b)^2{a_0^b-v_0^b\over (v_0^b)^2+(a_0^b)^2}
\simeq 1-2.2(s_L^b)^2-0.2(s_R^b)^2 .
\end{equation}
Hence, non-vanishing values for both
$s^b_R$ and $s^b_L$ have the effects of reducing
$\Gamma_b$, thus increasing the disagreement with the measurements.
Of course this behaviour is in part responsible for the drastic
improvement in the constraints on the $b$ mixing angles.
In addition, due to the  effect of the top mass
on the $Zbb$ vertex corrections, the constraint arising from $R_b$
slightly improves with larger $m_t$
(the measured $R_b$ favours a lower $m_t$ value).

In the case of the different leptonic asymmetries, the LEP experimental
values are not in complete agreement with the
assumption of universality, since $A^{FB}_\tau$ is somewhat larger
than $A^{FB}_{e,\mu}$. The very small SM value of the charged
lepton vector coupling
$v_0^l\simeq -0.036 $
implies that $A^0_l\simeq v^l/a^l$ is very sensitive to tiny
effects of new physics affecting $v^l$, as for example the shift
$\delta v^l=[(s_L^l)^2-(s_R^l)^2]/2$
induced by a mixing of the leptons. An
increase in $A^{FB}_\tau$ could then result from a non-zero
$s_R^\tau$. However, since this fermion mixing would modify
simultaneously the axial coupling $a^\tau$ by a similar amount, it is
easy to check that the constraints from $\Gamma_\tau$ do not allow the
50\% increase required to explain the measured $A_\tau^{FB}$
(in the presence of $s_R^\tau$,
$\delta A_\tau/A_\tau\simeq -2\delta \Gamma_\tau/\Gamma_\tau$).
New physics effects could be able to account
for these deviations only if they affect mainly
the $\tau$ vector coupling, while leaving the axial-vector
coupling close to its SM value.
Regarding the measurement of
$A_{LR}^{SLC}$, even if
one were to ignore the discrepancy with the
LEP measurement of $A^0_e$, the same type of argument would prevent
the possibility of explaining the measured value
by means of a mixing of the electron.

Clearly the deviations in $\Gamma_b$ and $A^{FB}_\tau$
 cannot be explained either by introducing a
$Z'$ boson of the type we have considered here, since these
new gauge interactions are universal and
would affect all generations. However, some
models involving a new gauge boson coupling mainly to the third
generation have been discussed in this context \cite{holdom94}.

In conclusion, LEP provides a powerful tool for the indirect search
of several types of physics beyond the SM.
Present observations do not hint to any of the new physics effects
that have been discussed here,
thus allowing for a
significant improvement of the limits on
the indirect effects induced by some of the new particles
that appear in many extensions of the SM.


\vskip 1.5 truecm

\begin{thebibliography}{99}
\frenchspacing

\bibitem{ll}
P. Langacker and D. London, \prd{38} (1988) 886.

\bibitem{fit}
E. Nardi, E. Roulet and D. Tommasini, \npb{386} (1992) 239.

\bibitem{mixlep}
E. Nardi and E. Roulet, \plb{248} (1990) 139; \\
G. Bhattacharyya \ea, \prl{64} (1990) 2870; \\
G. Bhattacharyya \ea, Mod. Phys. Lett. {\bf A6} (1991) 2921; \\
C.P. Burgess \ea,  \prd{49} (1994) 6115; \\
G. Bhattacharyya, \plb{331} (1994) 143.

\bibitem{fit6}
E. Nardi, E. Roulet and D. Tommasini, \prd{46} (1992) 3040.

\bibitem{zp-new}
P. Langacker and M. Luo, \prd{45} (1992) 278; \\
P. Langacker, M. Luo and A.K. Mann, Rev. Mod. Phys. 64 (1992) 87; \\
J. Layssac, F.M. Renard and C. Verzegnassi, \zpc{53} (1992) 97; \\
M.C. Gonzalez Garc\'\i a and J.W.F. Valle; \plb{259} (1991) 365; \\
G. Bhattacharyya \ea, Mod. Phys. Lett. {\bf A6} (1991) 2557; \\
O. Adriani \ea \plb{306} (1993) 187;\\
G. Altarelli \ea, \plb{263} (1991) 459; \\
G. Altarelli \ea, \plb{318} (1993) 139; \\
F. del Aguila \ea, \npb{372} (1992) 3. \par

\bibitem{cdf-mtop}
CDF Collaboration, F. Abe \ea, \prl{73} (1994) 225.

\bibitem{zfitter}
D. Bardin \ea, preprint CERN-TH 6443/92.

\bibitem{zp-paco}
For a review on $Z^\prime$ constraints, see F. del Aguila, M. Cveti\v c
and P. Langacker,
{\it  Proceedings of the 2nd International Workshop on
Physics and Experiments at Linear $e^+e^-$ Colliders},
Waikoloa, Hawaii, 1993, F. Harris \ea  eds.
(World Scientific, Singapore, 1993)
Vol. II, p. 490. \par


\bibitem{fc-limits}
D. London, in {\it
Precision Tests of the Standard Model}, ed.\ P. Langacker (World
Scientific, 1993); \
E. Nardi, in Ref. \cite{zp-paco},
Vol. II, p. 496;
C.P. Burgess \ea,  \cite{mixlep}. \par

\bibitem{fitnu} E. Nardi, E. Roulet and D. Tommasini,
{Phys. Lett.} {\bf B327} (1994) 319.

\bibitem{alr-slc}
SLD Collaboration, K. Abe \ea, \prl{73} (1994) 25.

\bibitem{lep94}
The LEP Electroweak Working Group, DELPHI 94-33 PHYS 364 (1994).\par

\bibitem{triumf92}
D.I. Britton \ea, \prl{68} (1992) 3000.

\bibitem{psi92}
G. Czapek \ea, \prl{70} (1992) 17.

\bibitem{roney}
J.M. Roney, in
{\it Proc. of the Second Workshop on Tau Lepton Physics},
ed. K.K. Gan (World Scientific, Singapore, 1993).

\bibitem{sirlin93}
A. Sirlin, in {\it Precision Test of the Standard Electroweak Model}
ed. \ P. Langacker  (World Scientific, Singapore, 1993).

\bibitem{pietrzyk}
B. Pietrzyk, Annecy preprint LAPP-EXP-94.07 (1994).

\bibitem{lang-luo}
See for example P. Langacker and M. Luo, in Ref. \cite{zp-new}.

\bibitem{holdom94}
B. Holdom, University of Toronto preprint UTPT-94-20 (1994).

\end{thebibliography}
\end{document}